\documentclass[aps,pra,twocolumn,notitlepage]{revtex4}
\usepackage{amsmath, amssymb}
\usepackage{bm}
\usepackage{natbib}
\usepackage{color}

\usepackage{graphicx}
\usepackage{float}

\begin{document}

\title{Quantum Time Crystals and Interacting Chiral Gauge Theories\\ in Atomic BECs III: Role of the Page-Wootters mechanism}

\author{P. \"Ohberg$^{1}$ and E. M. Wright$^{1,2}$}
\affiliation{$^1$SUPA, Institute of Photonics and Quantum Sciences, Heriot-Watt University, Edinburgh EH14 4AS, United Kingdom\\
$^2$James C. Wyant College of Optical Sciences, University of Arizona, \\ Tucson, Arizona 85721, USA}

\begin{abstract}  
We build the case that our published chiral soliton model ({\sl Phys. Rev. Lett.} {\bf 124}, 250402 (2019)) can lead to a genuine quantum time crystal for smaller atom numbers.  To do this we compare results from the chiral soliton model with exact numerical solutions for the ground state of the three-particle Schr\"odinger equation in the limit that the spatial profile of the soliton is largely independent of its center-of-mass motion.  The connection between the two approaches is found via the Page-Wootters mechanism, and this is what allows for dynamical evolution even in the quantum ground state, as required for a genuine quantum time crystal.
\end{abstract}

\maketitle

\section{Introduction}

Since Wilczek's seminal paper initiated the field of quantum time crystals in 2012 \cite{Wil12}, it has attracted considerable fundamental and applied interest.  His original model was composed of $N$ bosons with attractive interactions in a ring that was pierced by a flux tube, the key idea being that this model could yield a ground state composed of a soliton that was rotating around the ring.  This model was subsequently shown to always yield a non-rotating ground state \cite{Bruno}, but the time crystal idea had taken root.  Now there is active research in both continuous and discrete time crystals and for a variety of platforms including optics, ultracold atoms, metamaterials, spin systems, and both closed and open systems.  For recent reviews of the area see \cite{SacZak18,Sacha20,ZalLukMon23}.

A genuine time crystal involves a Hamiltonian system that exhibits sustained periodic motion even in its lowest-energy state. To the best of our knowledge there is only one acknowledged example of such a system, that involves long range interactions, but with no physical implementation to date \cite{KozKyr19}.  In a 2019 paper, hereafter referred to as QTC I \cite{OW19}, we proposed a model for a quantum time crystal based on a chiral soliton in a ring. In response Syrwid, Kosior, and Sacha (SKS) published a comment claiming that our system cannot realize a genuine time crystal \cite{SKS20}, and we responded to their comment \cite{OW_response}.  SKS have since published a paper with an extended version of their arguments \cite{SKS202}, but the issue has not been conclusively resolved.

The goal of this paper is to bolster the case that our original model in QTC I \cite{OW19} can indeed lead to a genuine quantum time crystal for smaller atom numbers, and in the limit that the spatial profile of the soliton is largely independent of its center-of-mass (COM) motion.  With these caveats the chiral soliton model constitutes a bona fide second example of a genuine quantum time crystal.  To do this we compared results from the chiral soliton model with exact numerical solutions for the ground state of the three-particle Schr\"odinger equation that were described in our second paper that we refer to as QTC II \cite{OW24}.  We also elucidate how the quantized COM of the soliton enters into the chiral soliton model, and this replaces our earlier approach \cite{OW_response}.  The connection between the results from the two approaches is made via the Page-Wootters mechanism \cite{PW83}, and this is what allows for dynamical evolution with repect to an internal clock even in the quantum ground state, as required for a genuine quantum time crystal.

\section{Chiral soliton model}

In this Section we discuss the basic chiral soliton model and its properties \cite{OW19}.  Of particular interest is the quantized COM motion of the soliton, and we expand and clarify our previous approach to this topic \cite{OW_response}.

\subsection{Basic equations}

We follow the notation of Ref. \cite{OW19} describing chiral solitons for a system of $N$ bosons on a ring of radius $R$.  The chiral NLSE is then $(W=0)$
\begin{equation}
i\hbar {\partial\psi\over \partial t} = \left [ -{\hbar^2\over 2m} {\partial^2\over \partial x^2}  -2a_1 N j(x) +gN|\psi|^2 \right ]\psi ,
\end{equation}
where $g$ is the parameter of the non-chiral contribution to the nonlinearity and $a_1$ the parameter for the chiral contribution.  This was derived employing the following nonlinear gauge transformation
\begin{equation}
\Psi(x,t) = \sqrt{N} \psi(x,t)e^{-i(\phi/2)+ (ia_1 N/\hbar)\int_{ }^x dx' |\psi(x',t)|^2},
\end{equation}
where the current nonlinearity is
\begin{equation}
j(x) = {\hbar\over 2mi} \left (\psi^*{\partial\psi\over \partial x} - \psi{\partial\psi^*\over \partial x}
\right )  .
\end{equation}
Here $\phi = -qx/R$, $q$ being the winding number of the applied LG laser field.  The chiral soliton solution for this model in a frame rotating at velocity $u$ is given in Eq. (15) of QTC I \cite{OW19}, or Eq. (18) in scaled units.  Note that the soliton centroid is chosen at $x=\theta=0$ at $t=0$ in these solutions, where $\theta=x/R$ is the angular position around the ring.

\subsection{Conserved quantities}

We can find the conserved quantities associated with this chiral soliton model using the results from the paper by Jackiw \cite{Jackiw}: For this we note that the chiral parameters are related via $\lambda/2 \rightarrow a_1/\hbar$, and here the wave function is normalized to unity whereas for Jackiw it is normalized to $N$.  Then according to Jackiw's paper, and adapted to our notation, the energy in the lab frame for the N-particle system is
\begin{equation}
E_{LAB}= N \int_0^{C} dx~\left [ {\hbar^2\over 2m} \left |{\partial\psi\over\partial x}\right |^2 + {gN\over 2} |\psi|^4 \right ] ,
\end{equation}
the integral being taken over the ring circumference $C=2\pi R$, and the COM momentum density is
\begin{equation}\label{Pdens}
{\cal P} = {N\hbar q\over 2R}|\psi|^2 + mNj + a_1N |\psi|^4  ,
\end{equation}
with the first term on the right-hand-side of the momentum density arising from the gauge potential induced by the LG beam.  Then both the energy and COM momentum are conserved
\begin{equation}
{dE_{LAB}\over dt} = {d\over dt} \int_0^{C} dx~{\cal P} = 0.
\end{equation}

\subsection{Quantized COM velocity}

We next consider the rotating chiral solitons in Eq. (15) of QTC I \cite{OW19} and evaluate their COM velocity. To do this we recognize that the quantities ${\cal P}/|\psi|^2$ and $j/|\psi|^2$ are the local values of the momentum and velocity, respectively, both of which may vary around the ring. Then using Eq. (\ref{Pdens}) and defining the COM momentum and velocity averaged over the ring
\begin{equation}
{\mathsf P} =\int_0^{C} dx~|\psi|^2 \left ( {{\cal P}\over |\psi|^2} \right )
\quad u = \int_0^{C} dx~ |\psi|^2 \left ({j\over|\psi|^2} \right ),
\end{equation}
we obtain
\begin{equation}
{\mathsf P} ={N\hbar q\over 2R} + M u + a_1N^2\int_0^{C} dx~|\psi|^4,
\end{equation}
where $M=mN$ is the total mass of the rotating soliton and we used the fact that the wave function is normalized over the ring.  Note that the COM momentum so defined has both a kinematical term involving the soliton velocity $u$ as well as a dynamical contribution proportional to the parameter $a_1$ which characterizes the strength of the chiral nonlinearity.

In the thermodynamic limit of the mean-field theory the COM momentum ${\mathsf P}$ may be treated as a continuous variable for which quantization is not relevant.  On the other hand, for a mesoscopic system of bosons the quantized COM motion may be incorporated into the analysis by appealing to the Sommerfeld quantization condition
\begin{equation}
\int_0^{C} {\mathsf P}~dx = 2\pi p \hbar  = N\pi\hbar q +2\pi R M u + a_1N^2\Gamma ,
\end{equation}
where $p=0,\pm 1,\pm 2,\ldots$ is an integer, and $\Gamma$ is a dimensionless factor that accounts for the inhomogeneous spatial profile of the chiral soliton ($\Gamma=1$ for a homogeneous profile) with
\begin{equation}
\Gamma =2\pi R \int_0^{C} dx~|\psi|^4 .
\end{equation}
This result may be rearranged as
\begin{equation}
{2MuR\over \hbar} = 2\left [ p - N\left ({q\over 2} \right ) \right ] - {a_1 N^2\Gamma\over\pi\hbar} , \quad p=0,\pm 1,\pm 2,\ldots
\end{equation}
The idea is then to choose $p=p_{min}$ to find the minimum speed $|u|$ of the chiral soliton in the ground state.  The transition to the scaled units used in QTC I \cite{OW19} is accomplished using ${2muR\over\hbar} \rightarrow u$, giving
\begin{equation}\label{uEq}
u = {2\over N} \left [ p - N\left ({q\over 2} \right ) \right ] - {a \Gamma\over\pi } , \quad p=0,\pm 1,\pm 2,\ldots
\end{equation}
where $a=a_1N/\hbar$.  This result replaces our previous expression for the quantized velocity \cite{OW_response}.  A couple of limiting cases highlight the upshot of this result:

\begin{itemize}

\item {\bf Thermodynamic limit:}  In the thermodynamic limit $N\rightarrow \infty$ the quantity $p/N$ approaches a continuous variable whose value can always be chosen to cancel the combined terms ${q\over 2} + {a \Gamma\over 2\pi }$, giving $u\rightarrow 0$.  In this case a genuine time crystal is not possible.  This is the argument used by SKS for the case of the Wilczek model \cite{SKS202}.

\item {\bf Few-particle limit:}  For a few particles a genuine time crystal is possible according to this model incorporating the Sommerfeld quantization condition.  In particular, if $|{a \Gamma\over\pi }| \ll1$ and $q$ is even, then the velocity with smallest magnitude is $u=u_{dyn}=- {a \Gamma\over\pi }$ occurring  when $p=p_{min}=N\left ({q\over 2} \right )$.  This gives a genuine time crystal as the COM velocity $u$ need not equal zero in the ground state, but rather it is given by the dynamical contribution $u_{dyn}$ to the velocity proportional to the chiral nonlinear parameter.

\end{itemize}
We note that although we find a velocity for the centroid of the chiral soliton the theory gives no indication of the position of the centroid of the soliton at a chosen initial time, all possible values of the position being equally valid.  This reflects the fact that although the chiral nonlinearity biases one direction of rotation over the other it does not provide an absolute origin in either position or time. This is akin to a mode-locked laser in which one knows it will emit a periodic pulse train, but the arrival time of the first pulse at the output coupler is subject to variation from realization to realization after the laser is turned on.

\begin{figure}
\includegraphics[width=0.8\linewidth,clip]{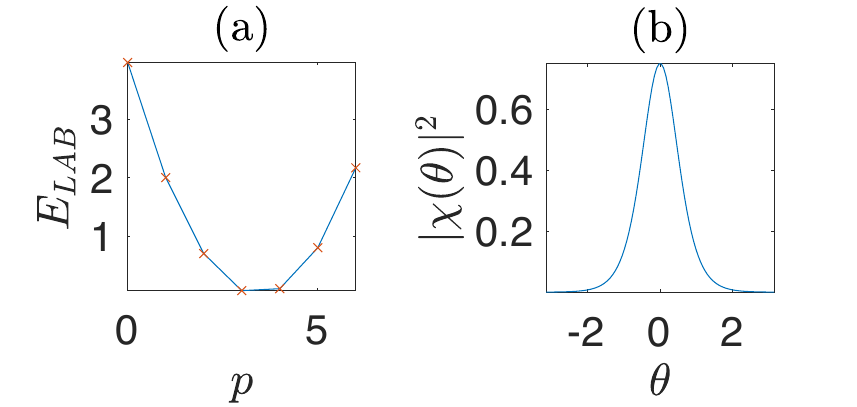}
\caption{Sample results for $q=2, g=-6, a=0.1$ and $N=3$.  In particular, plot (a) shows $E_{LAB}$ versus $p$ with a minimum at $p_{min}=N(q/2)$, and plot (b) shows the chiral soliton probability density $|\chi(\theta)|^2$ versus the angular position $\theta$.}\label{Fig1}
\end{figure}

\begin{figure}
\includegraphics[width=1.0\linewidth,clip]{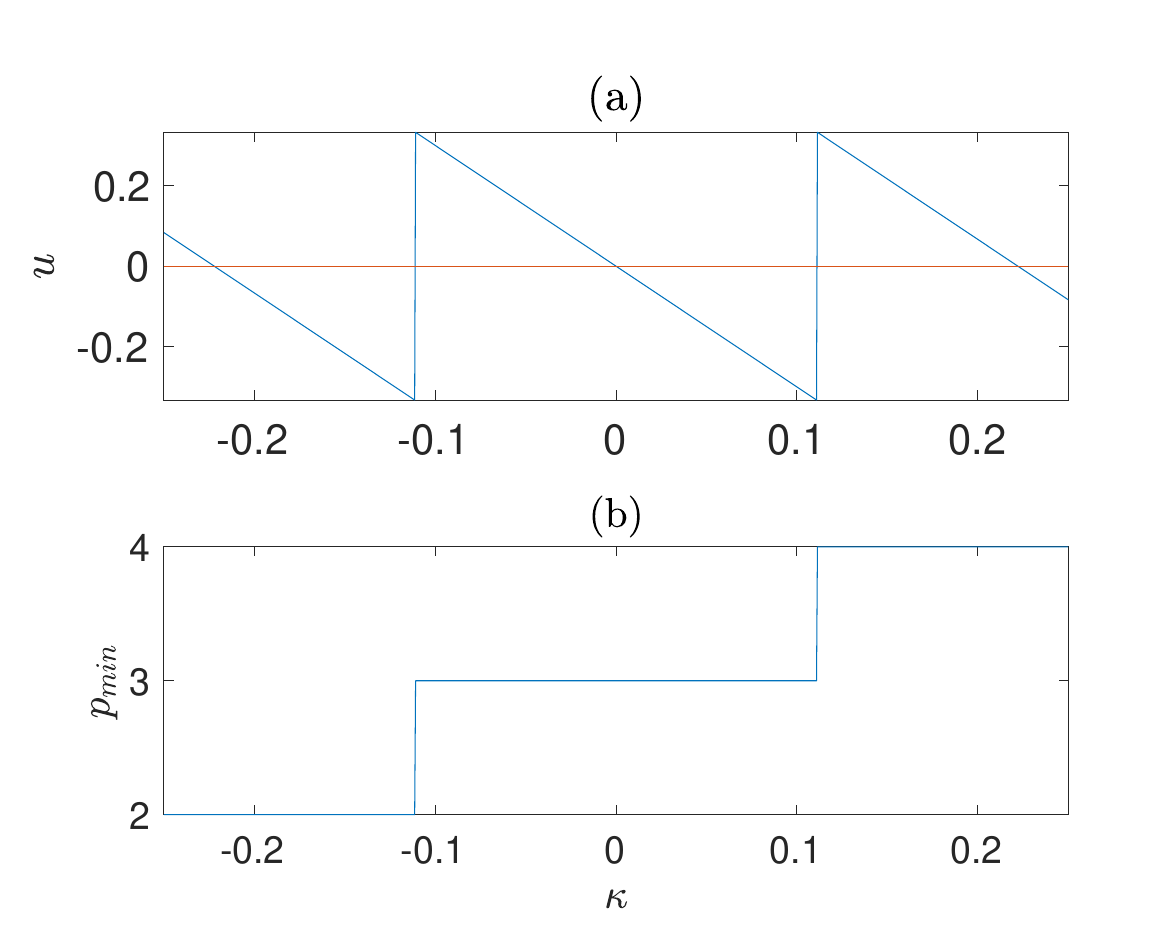}
\caption{For $q=2,g=-6$, and $N=3$ we show (a) the soliton velocity $u$, and (b) $p_{min}$ versus $\kappa$.}\label{Fig2}
\end{figure}

\subsection{Ground state energy}

The idea here is to employ the approximate chiral soliton in Eq. (15) from QTC I \cite{OW19} to evaluate the ground state energy.  SKS evaluated the conserved energy in the lab frame using this approach \cite{SKS20} with the result
\begin{equation}
E_{LAB} = N{\cal E}_{LAB} = -{mg^2N^3\over 24\hbar^2} + \left ( 1 + {a^2\over 3}\right ) {Nmu^2\over 2}  ,
\end{equation}
where the dimensionless parameter $a=Na_1/\hbar$.  The first term on the right-hand-side is a constant that we hereafter subsume into $E_{Lab}$.  The chiral soliton solution depends on the nonlinear parameter  $(g-2au)$ in scaled units.  For our treatment we assume that $|g| \gg |2au|\sim a^2$, and the normalized chiral soliton profile may be approximated as
\begin{equation}\label{chi0}
\chi(\theta) = {1\over \sqrt{2b}} {1\over \cosh(\theta/b)}  , \quad b = {4\over |g|}, 
\end{equation}
where $\theta=x/R=[-\pi,\pi]$ is the angular position around the ring.  In this limit that we adopt here the spatial profile of the soliton is dominated by the non-chiral portion of the nonlinearity via $g$, but as we shall see the soliton velocity can still depend on the chiral contribution proportional to $a$.  This approach involves neglecting terms of order $a^2$, so the soliton energy in the lab frame and in scaled units may be approximated as
\begin{equation}
E_{LAB} = {Nu^2\over 4}  .
\end{equation}
Substituting the expression for the velocity in Eq. (\ref{uEq}) we finally obtain
\begin{equation}
E_{LAB} = {\left ( \left [ p - N\left ({q\over 2} \right ) \right ] - {a \Gamma\over\pi } \right )^2 \over N},
\end{equation}
where within our chiral soliton approximation $\Gamma={\pi|g|\over 6}$.  The ground state energy is found by choosing $p=p_{min}$ that minimizes the energy above.  We also note that the soliton velocity may be found from 
\begin{equation}
u = {\partial E_{LAB}\over\partial p}  ,
\end{equation}
evaluated for $p=p_{min}$, so that $u$ is the group velocity of the chiral soliton as expected.  We note that this treatment of the COM motion is analogous to that used by SKS \cite{SKS202} in their treatment of the Wilczek model with a magnetic-like flux $\alpha$.  In their case $\alpha$ was due to an externally imposed flux-tube whereas in our case the effective flux is $\alpha={a \Gamma\over\pi }$ that is explicitly due to the chiral nonlinearity.

Figure \ref{Fig1} shows sample results for this using $q=2, g=-6, a=0.3$, and $N=3$. (The small particle number is taken to compare against the few-particle solutions later).  In particular, plot (a) shows $E_{LAB}$ versus $p$ with a minimum at $p_{min}=N(q/2)$, and plot (b) shows the chiral soliton probability density.  As parameters are varied both the value of $p_{min}$ and $u$ vary.  Figure \ref{Fig2} shows both (a) the soliton velocity $u$, and (b) $p_{min}$ versus $\kappa$, where $a=N\kappa$. (We vary the parameter $\kappa$ to connect with our few-particle simulations).
\begin{figure}
\includegraphics[width=0.9\linewidth,clip]{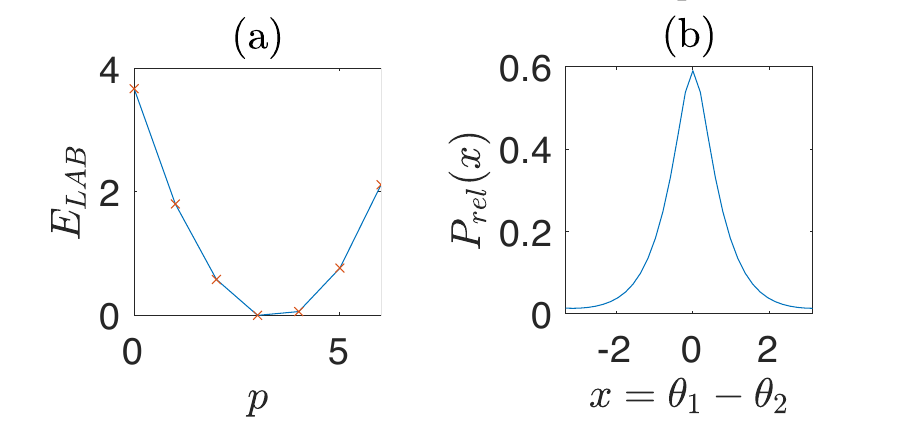}
\caption{Sample results for $q=2, g'=-4$, and $\kappa=0.1$.  In particular, plot (a) shows $E_{LAB}$ versus $p$ with a minimum at $p_{min}=N(q/2)=3$, and plot (b) shows the probability density $P_{rel}(x)$ in Eq. (\ref{Prel}) as a function of the relative coordinate $x=(\theta_1-\theta_2)$.}\label{Fig3}
\end{figure} 
Figure \ref{Fig2}(a) reveals some generic features for $q$ even and in the parameter region of interest:

\begin{itemize}

\item For $a=\kappa=0$ the chiral soliton velocity $u=0$.  This verifies the fact that the non-chiral Wilczek model does not give a genuine time crystal.

\item As $\kappa$ is varied the attainable speed has an upper limit of $|u|\sim 1/N$, so in the thermodynamic limit genuine time crystal behavior is not possible as noted before.

\item Around the origin the chiral soliton velocity scales with $\kappa$ and need not vanish.

\end{itemize}
These features again highlight that, according to the chiral soliton model, the ground state of our system for smaller $N$ is a genuine time crystal for which the ground state reflects a state of motion of the chiral soliton, that is, a localized and rotating solution.  Similar results to Fig. \ref{Fig2} are obtained for $q$ odd except $\kappa=0$ is translated horizontally to one of the vertical transition points.  In the remainder of this paper we choose $q=2$ for illustration.

\subsection{Three-particle Schr\"odinger equation}

The analytic chiral soliton model is interesting but can be challenged on the basis that using the Sommerfeld quantization condition to incorporate quantized COM motion into the model is ad hoc. A more fundamentally based approach is needed to validate the approximations used.  In this Section we build the case for this by solving numerically for the ground state of the three-particle Schr\"odinger equation and comparing against the chiral soliton model results - even for such a small particle number we find impressive agreement.

Briefly, we used the approach from QTC II \cite{OW24} to generate results for the quantum ground state for a variety of parameters with $N=3$ that correspond to the results in the previous Section.  For this case the density for the three bosonic particles may be written as
\begin{equation}
\rho(\theta_1,\theta_2,\theta_3) \equiv \eta(\theta_1-\theta_2)+\eta(\theta_1-\theta_3)+\eta(\theta_2-\theta_3) ,
\end{equation}
which by construction is symmetric under exchange of any pair of particle coordinates.  The Schr\"odinger equation for $\Psi(\theta_1,\theta_2,\theta_3,t)$ may then be written as
\begin{eqnarray}
i{\partial\Psi\over\partial t} &=& \sum_{j=1}^3  \left ( -i{\partial\over \partial \theta_j}- {q\over 2} -\kappa\rho(\theta_1,\theta_2,\theta_3)) \right )^2 \Psi  \nonumber\\&& +{g'\over 2} \rho(\theta_1,\theta_2,\theta_3) \Psi .
\end{eqnarray}
To map to the chiral soliton results the parameters are related using
\begin{equation}
g = {Ng'\over 2}, \quad a = N\kappa.
\end{equation}
In scaled units the ground-state wave function is written as
\begin{equation}
\Psi(x,y,s,t) = e^{-iE_{LAB}t + ip s} \varphi(x,y),
\end{equation}
where $s={1\over 3}(\theta_1 + \theta_2 +\theta_3), x=(\theta_1-\theta_2)$ and $y=(\theta_1+\theta_2-2\theta)$ are Jacobi coordinates for the three-particle problem.  In particular, $s$ is the COM position and $x$ is the relative position of two-particles, the probability density with respect to the relative coordinate being given by
\begin{equation}\label{Prel}
P_{rel}(x) = \int_{ }^{ } dy~|\varphi(x,y)|^2.
\end{equation}
We note that the corresponding probability density $P_{COM}(s)$ in terms of the COM coordinate $s$ is uniform, which means that the full quantum ground state does not explicitly reveal any soliton localization and accompanying rotation though these do appear explicitly in the chiral soliton model.  We shall return to this point in the next Section.

\begin{figure}
\includegraphics[width=1.0\linewidth,clip]{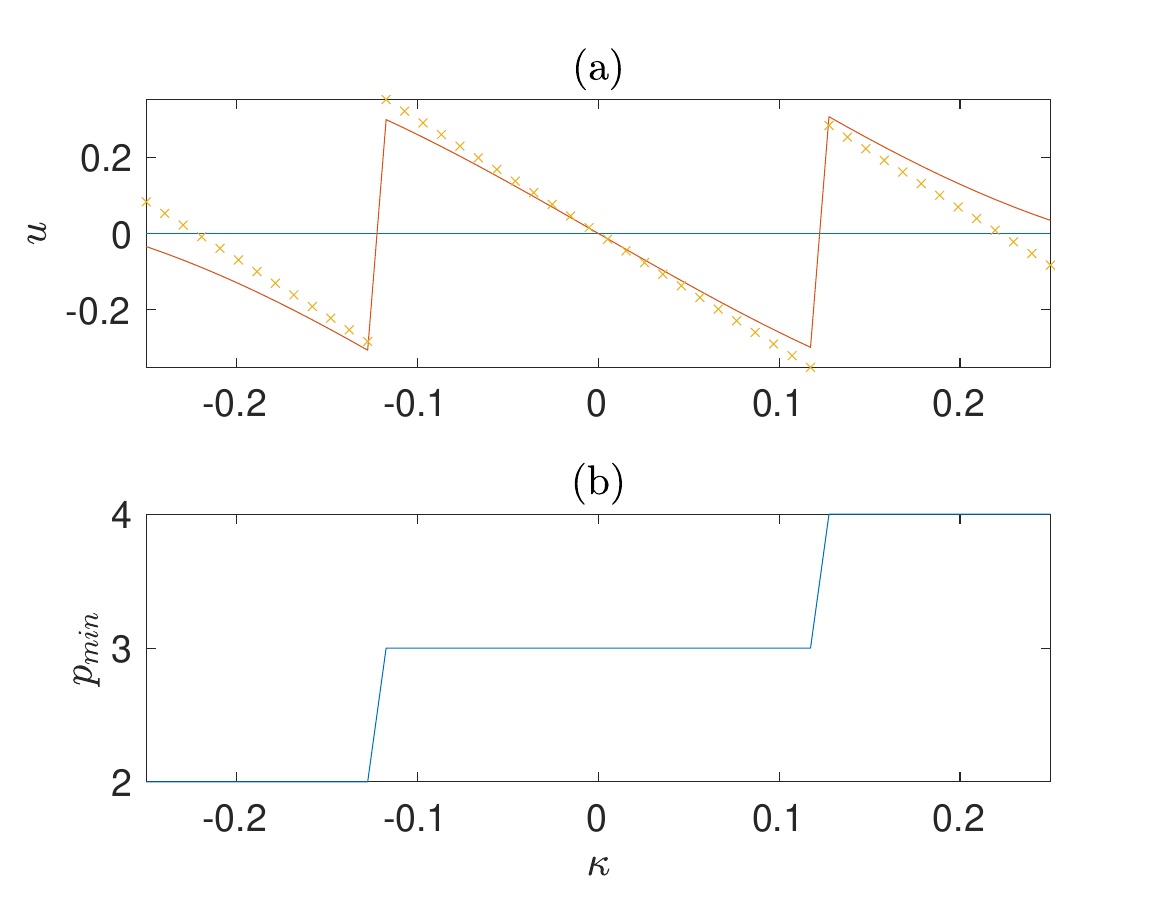}
\caption{(a) The group velocity $u={\partial E_{LAB}\over \partial p}$, and (b) $p_{min}$ versus $\kappa$.  In particular, in plot (a) the solid line is the result of the full quantum simulation and the crosses are the results from the chiral soliton theory.}\label{Fig4}
\end{figure}

Figure \ref{Fig3} shows sample results using $q=2, g'=-4$, and $\kappa=0.1$ ($g=-6, a=0.3)$.  In particular, plot (a) shows $E_{LAB}$ versus $p$ with a minimum at $p_{min}=N(q/2)=3$, and plot (b) shows the probability density $P_{rel}(x)$.  The close agreement between the results in Figs. \ref{Fig1} and \ref{Fig3} is clear.  As parameters are varied both the value of $p_{min}$ and the group velocity $u$ vary.  Figure \ref{Fig4} shows both (a) the group velocity $u$, calculated numerically using a three or five-point approximation from the $E_{LAB}$ versus $p$ plot in Fig. \ref{Fig3}(a), and (b) $p_{min}$ versus $\kappa$ with all other parameters fixed.  In particular, in plot (a) the solid line is the result of the full quantum simulation and the crosses are the results from the chiral soliton theory - they agree very well for smaller $\kappa$ but one can see the deviations for larger $\kappa$ (or $a$) as expected.  Again the close agreement between the results in Figs. \ref{Fig2} and \ref{Fig4} are clear.

The close agreement between the results from the chiral soliton model and the three-particle Schr\"odinger equation can be taken as validation of our use of the Sommerfeld quantization condition within the limits considered. We observe similar agreement between the results from the analytic chiral soliton theory and the full quantum ground state for $N=3$ over a range of parameters within the limitations considered.  In particular, we have assumed that in the chiral soliton theory $|g|\gg1, a^2 \ll1$ and the spatial profile of the soliton is determined dominantly by the non-chiral part of the nonlinearity, and is largely independent of the COM motion of the soliton.  Our main interest here is the region $\kappa \ll 1$ where $p_{min}=N(q/2)$ and the soliton velocity is given by the dynamical contribution $u = u_{dyn}=-{N\kappa\Gamma\over \pi}$, the linear region in Figs. \ref{Fig2} and \ref{Fig4}.  Finally we note that the good agreement found here for $N=3$ suggests that the agreement will be even better for larger $N$, so that the chiral soliton model will be of more general utility than presented here.

\section{Role of the Page-Wootters mechanism}

It appears that we have now reached an impasse in which the chiral soliton model presents a localized and rotating soliton solution, albeit whose initial centroid at $t=0$ can be chosen uniformly over the ring, whereas the full quantum solution based on the ground state of the three-particle Schr\"odinger equation presents no such localization and rotation.  On the other hand, the ground state energy from the two approaches agree very well, and the soliton velocity from the chiral soliton model agrees very well with the group velocity obtained from the full quantum solution.

The resolution of this impasse may be found by appealing to the Page-Wootters mechanism \cite{PW83} that is described in their aptly titled paper "Evolution without evolution: Dynamics described by stationary observables."  According to this mechanism a system in its ground state can display dynamical evolution of a subsystem of interest due to correlations with an internal quantum clock, the subsystem plus clock comprising a closed quantum system.  Page and Wootters advanced this mechanism in the context of the ground state solution of the Wheeler-DeWitt equation for the universe in order to demonstrate that being in the ground state does not discount dynamical evolution \cite{PW83}.

For our use the system is the chiral soliton, and the internal clock is associated with the quantum COM motion of the soliton.  There are three assumptions required for the applicability of the Page-Wootters mechanism:
\begin{enumerate}

\item The internal clock does not interact with the system.  This is true under the conditions assumed here, $|g|\gg1, a^2\ll1$, for which the spatial profile of the chiral soliton is largely independent of the COM velocity $u$, and so the chiral soliton and COM motion are not directly coupled.

\item The internal clock is entangled with the system.  This follows from the fact that the ground state from the three-particle Schr\"odinger equation may be viewed as an entangled state of chiral solitons with differing COM positions uniformly distributed over the ring, and this is why the full quantum results do not show explicit localization or COM motion.  Appendix A provides more details for the interested reader.

\item The combined internal clock and system are in an energy eigenstate of the closed system.  This is certainly the case for our problem for which the energy and Hamiltonian are written in the lab frame which incorporates both many-body interactions and COM variables.

\end{enumerate}
We may therefore appeal to the Page-Wootters mechanism to resolve the above impasse: Even though the full quantum solution based on the three-particle Schr\"odinger equation shows no signs of localization or COM motion, dynamics with respect to the internal clock supplied by the COM motion of an initially localized solution of the chiral soliton model (we chose $x=x_0$ for the soliton center at $t=0$) is a consistent solution given the applicability of the Page-Wootters mechanism to our problem.  Based on this analysis we contend that our chiral soliton model gives rise to a genuine quantum time crystal for smaller particle numbers $N$, although we concur that a genuine time crystal will not be possible in the thermodynamic limit $N\rightarrow \infty$ \cite{SKS20}.

\section{Time crystal behavior from weak measurement}

Our analysis leading to the conclusion that we have a genuine quantum time crystal did not involve an external measurement or clock that monitors the COM motion of the chiral soliton, and as such it leaves undetermined the specific value of the centroid of the soliton at an initial time $t_0$.  In a previous paper QTC II \cite{OW24} we presented results for $N=2,3$ showing that quantum time crystal behavior can be initiated via weak position measurements, the weak measurement at $t=t_0=0$ establishing both the initial time and the initial value of the centroid of the soliton.  An open question is then whether such a weak external measurement will yield the same soliton velocity as the dynamical value $u_{dyn}$ from the chiral soliton theory, and the answer is generally no.  This is not surprising  as the Page-Wootters mechanism involves an internal clock and a closed system whereas an external measurement renders the system open.  That is, the action of even a weak position measurement, that involves the removal of a particle, can modify the momentum of the remaining particles, giving rise to an excited state.  We present some simulations of this here for $N=3$.

\begin{figure}
\includegraphics[width=1.0\linewidth,clip]{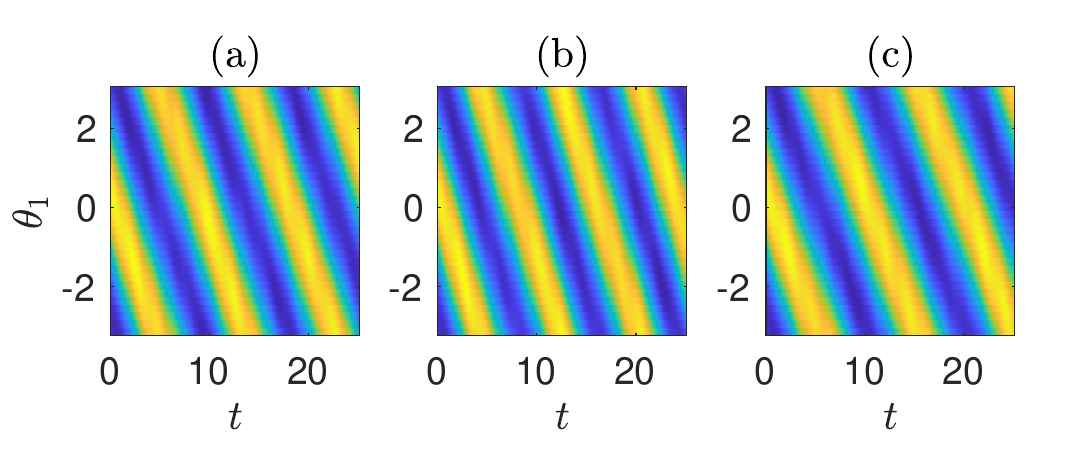}
\caption{Simulations for $N=3$ and $g'=-4$ of the single-particle probability density $p(\theta_1,t)$ following a weak measurement of particle $3$. (a) $\kappa=0$, (b) $\kappa=0.1$, and (c) $\kappa=-0.1$.}\label{Fig5}
\end{figure} 

The procedure is as follows \cite{OW24}: Denoting the ground state wavefunction for $N=3$ as $\Psi(\theta_1,\theta_2,\theta_3)$ and assuming that the position of particle $3$ is measured yielding a value $\theta_3^{(0)}$, but allowing for the measurement to be imprecise, then the wave function for the two remaining particles may be written as
\begin{eqnarray}\label{psi_3init}
    \psi(\theta_1,\theta_2,t\!=\!0)&=&\Psi(\theta_1,\theta_2,\theta_3^{(0)})\nonumber\\&=&\!\! \int_{-\pi}^\pi \!\! \!\! \!\! d\theta_3 G(\theta_3-\theta_3^{(0)})\Psi(\theta_1,\theta_2,\theta_3) ,
\end{eqnarray}
where the function $G(\theta_3-\theta_3^{(0)})$ reflects the uncertainty of the position measurement.  After the measurement the wave function $\psi(\theta_1,\theta_2,t=0)$ (suitably normalized) for the remaining particles is given by Eq. (\ref{psi_3init}), and subsequent time development is governed by the two-particle Schr\"odinger equation
\begin{eqnarray}\label{EqNeq2}
i{\partial\psi\over\partial t} &=& -\left ( {\partial^2\over\partial \theta_1^2} +  {\partial^2\over\partial \theta_2^2} \right )\psi \nonumber\\&&+2i \left ( {q\over 2} +\kappa\rho \right )\left ({\partial\over\partial \theta_1} + {\partial\over\partial \theta_2} \right )\psi \nonumber\\&& + 2 \left ( {q\over 2} +\kappa\rho \right )^2\psi + {g\over 2}\rho \psi, 
\end{eqnarray}
where $\rho=\eta(\theta_1-\theta_2)$ is the scaled density for the two remaining particles.  From this two-particle wave function we calculate the single-particle probability density for measuring a second particle
\begin{equation}
    p(\theta_1,t) = \int_{-\pi}^\pi d\theta_2 |\psi(\theta_1,\theta_2,t)|^2 .
\end{equation}
For the simulations to be presented we set
\begin{equation}
G(\theta-\theta^{(0)})={\cal N} \cos^2 \left ( {\theta-\theta^{(0)} \over 2} \right )
={{\cal N}\over 2} \left [ 1 +  {e^{i\theta} + e^{-i\theta}\over 2} \right ],
\end{equation}
where ${\cal N}$ is a normalization constant, and we used $\theta^{(0)}=0$ without loss of generality.  From this form of $G(\theta-\theta^{(0)})$ we see that the weak position measurement can alter the momentum of the system by $0,\pm 1$ in scaled units, giving changes in velocity of $0,\pm 0.333$ for $N=3$.  This means that the measurement by necessity creates an excited state, and it is already known that time crystal behavior can arise for excited states \cite{SKS17}.  Thus, any velocities arising from quantum dynamics following the weak measurement can differ from those from the chiral soliton theory since the measurement, albeit weak, can provide an impulse to the momentum of the chiral soliton.

Some illustrative examples are shown in Fig. \ref{Fig5} for the single-particle probability $p(\theta_1,t)$ for $N=3, g'=-4$, and $\kappa=0,\pm 0.1$.  The case with $\kappa=0$ corresponds to the Wilczek model with no chiral contribution, and inspection of Fig. \ref{Fig5}(a) reveals that time crystal behavior appears after the weak measurement that is applied at $t=0$, and persists for far longer times than shown, with a soliton velocity $u_{\kappa=0}\sim -0.7$.  Thus, although a genuine time crystal is not possible for the Wilczek model, the weak measurement can initiate time crystal behavior \cite{OW24}, the sense of soliton rotation being dependent on the orbital angular momentum $q$ of the applied laser beam. Figure \ref{Fig5}(b) shows $p(\theta_1,t)$ for $\kappa=0.1$, and inspection of the data reveals $u_{0.1}\sim -0.8$, so the soliton speed is increased with respect to $\kappa=0$.  We may intuit this by assuming that the soliton velocity for $\kappa \ne 0$ is a combination of $u_{\kappa=0}$ and the dynamical value $u_{dyn}=-N\kappa\Gamma/\pi$. Then for $\kappa > 0$ both terms are negative so the soliton speed is increased with respect to $\kappa=0$. On the other hand, Fig. \ref{Fig5}(c) shows the plot for $\kappa=-0.1$, and inspection of the data reveals $u_{-0.1}\sim -0.65$, and the soliton speed is decreased with respect to $\kappa=0$. (Beyond this particular example, the key to this argument is the relative sign between the two contributions to the velocity).  Although qualitative this discussion does highlight that the soliton velocity depends on $\kappa$ in a way that reveals signatures of there being a contribution from the chiral nonlinearity. The fact that the velocities obtained from the data are asymmetric with respect to the case with $\kappa=0$ is perhaps not surprising as the chiral model is not Galilean-invariant, so transforming from the lab frame to one rotating at the velocity for $\kappa=0$ does not imply the symmetry assumed around $\kappa=0$ for the chiral soliton model.

With these sample results we hope to illustrate that the analysis of the data from quantum time crystal behavior initiated by weak measurements need not coincide with the chiral soliton theory and that this is not surprising. The weak measurement data is with respect to an external clock and the chiral soliton theory is with respect to an internal clock.  On the other hand, it is important to note that from a practical perspective that quantum time crystal behavior can survive and be initiated by the measurement process, even for the Wilczek model with no chiral nonlinearity and that does not constitute a genuine time crystal.  In addition, we have here used a position measurement that extracts a particle from the system, and perhaps there is a less destructive type of measurement, eg. interferometric, that could more closely correlate with the chiral soliton theory.

\section{Summary and conclusions}

In this paper we have made the case that our original model in QTC I \cite{OW19} does indeed lead to a genuine quantum time crystal for smaller atom numbers, and in the limit that the spatial profile of the soliton is largely independent its COM motion.  By comparing results from the chiral soliton model with those of the exact ground state of the three-particle Schr\"odinger equation, we found excellent agreement for the three-particle lab frame energy and velocity, although the chiral soliton model involves the dynamics of a localized and moving soliton whilst the full quantum theory shows no such features.  The connection between the the two approaches is made via the Page-Wootters mechanism \cite{PW83}, and this is what allows for dynamical evolution with respect to an internal clock even in the full quantum ground state, as required for a genuine time crystal.  The chiral soliton model therefore constitutes a second bona fide example of a genuine quantum time crystal.  The limitation to small particle numbers $N$ is clearly a barrier to comparison with current experiments, and future work will deal with the extension of the approach to larger mesoscopic systems.  On the other hand, perhaps future advances and technology will allow for the realization of such truly microscopic quantum time crystals that can be initiated by the measurement process. Only time will tell.

\appendix 

\section{Entangled ground state}

The goal of this Appendix is to build the many-body entangled ground state for a system of N identical bosons starting from the results of the chiral soliton theory, and thereby show how the Page-Wooters mechanism enters our proposal \cite{PW83}.  The normalized chiral soliton $\chi(\theta)$ centered at $\theta=0$ is given in Eq. (\ref{chi0}) of the main text, and we denote by $\chi_s(\theta)=\chi(\theta-s)$ the soliton translated by an angular displacement $s$. Then the many-particle wave function for the chiral soliton subsystem may be written in the Hartree approximation as
\begin{equation}
\Psi_{s}(\theta_1,\theta_2,\ldots\theta_N,t) =  \prod_{j=1}^N \chi(\theta_j-s) .
\end{equation}
We note that there is an implied time dependence in this wavefunction as the COM position of the localized soliton rotates with time according to $s(t)=s_0+ut$, with $u$ the soliton group velocity.  (For simplicity in presentation we leave out the exponential time-dependence associated with the ground state energy $E_{LAB}$).  The COM coordinate $s$ therefore represents the internal clock subsystem, and we recall that in the current scenario the spatial extent of the soliton subsystem is assumed to be largely independent of the COM motion.

To proceed we write the N-particle state above in second-quantized form as
\begin{equation}
|\Psi_s\rangle_{rel} = {1\over\sqrt{N!}} \prod_{j=1}^N \left [ \int_0^C d\theta_j~\chi(\theta_j-s)\hat\psi^\dagger(\theta_j) \right ] |0\rangle ,
\end{equation}
and this represents the internal or relative state describing the translated soliton subsystem.  Then the many-body state for a single localized soliton located at COM position $s$ can be expressed as the tensor product $|s\rangle_{COM} \otimes |\Psi_s\rangle_{rel}$, where $|s\rangle_{COM}$ is an angular position COM state peaked at $s$ that represents the quantum state of the internal clock subsystem.  

In the final step we recognize that the position localized N-particle soliton states $|s\rangle_{COM} \otimes |\Psi_s\rangle_{rel}$ for each $s$ do not reflect the required full translational invariance of the exact ground state on the ring.  There are two ways to restore the full translational invariance: First, we can form a superposition of all the positions $s$ around the ring to form a COM momentum eigenstate
\begin{equation}\label{A3}
|\Psi_p(t) \rangle = {1\over \sqrt{2\pi}} \int_0^C ds~e^{ips} |s\rangle_{COM} \otimes |\Psi_s\rangle_{rel} ,
\end{equation}
with $p$ the eigenvalue of the COM momentum, the factor $e^{ips}$ being included since the momentum is the generator of COM translations.  The second equivalent way to view this is that the momentum eigenstates on the ring can be formed as Fourier transforms of position localized states.

The quantum ground state in Eq. (\ref{A3}) is achieved by setting $p=p_{min}$ and this provides an approximation to the full quantum ground state of the many-body system built up from the localized and rotating solutions from the chiral soliton theory, and this provides the relation to the Page-Wootters mechanism \cite{PW83}.  In contrast, by virtue of the sum over all COM positions the quantum state in Eq (\ref{A3}) has no signatures of localization or rotation.  In addition, the quantum state is clearly an entangled state as it cannot be factorized due to the fact that both $|s\rangle$ and $|\Psi_s\rangle_{rel}$ depend on $s$ and the integrand cannot be simplified.

In this Appendix we have treated the general N-particle case and it is clear that the entanglement of the ground state survives even in the thermodynamic limit $N\rightarrow \infty$.  The conditions for the Page-Wootters mechanism to apply also extends to the thermodynamic limit, and this begs the question of why a genuine time crystal does not survive in the thermodynamic limit?  The reason is not the failure of the Page-Wootters mechanism but the fact that the soliton group velocity is zero in the thermodynamic limit, so there is no COM motion.

\end{document}